# Spatial Audio Rendering for Real-Time Speech Translation in Virtual Meetings

Margarita Geleta[1,*], Hong Sodoma[2], Hannes Gamper[2],

[1] Berkeley AI Research Laboratory, Berkeley, CA, USA
[2] Microsoft Corporation, Redmond, WA, USA
[*] Research conducted during an internship at Microsoft Corporation.

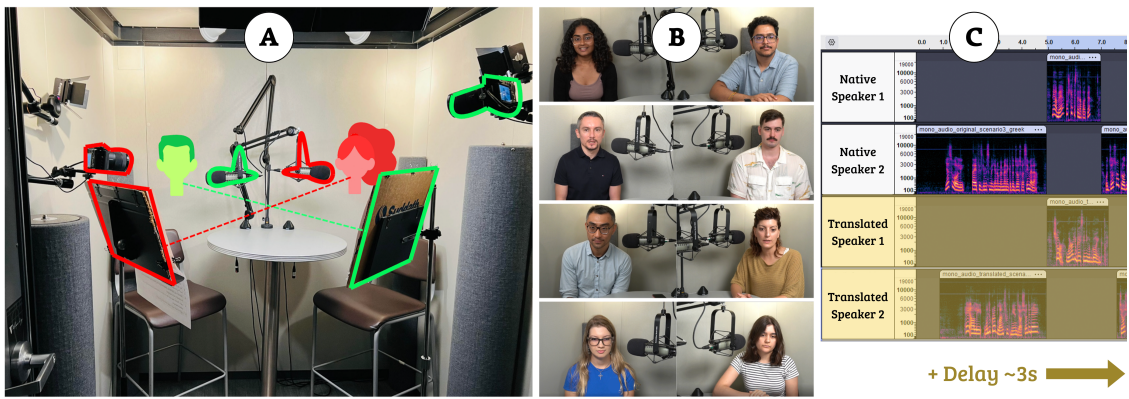

**Figure 1.** Overview of our WoZ setup and processing pipeline: (A) Recording booth with two bilingual confederates, each captured with a dedicated camera, microphone, and cardboards displaying scripted dialogues; (B) Confederate pairs across the four studied languages, showing both same-gender and mixed-gender pairs; (C) Audio processing workflow: native speech channels (top) are recorded cleanly and paired with English translations voiced by the same confederates (bottom). A controlled translation delay is introduced to simulated realistic machine translation latency.

## Abstract

Language barriers in virtual meetings remain a persistent challenge to global collaboration. Real-time translation offers promise, yet current integrations often neglect perceptual cues. This study investigates how spatial audio rendering of translated speech influences comprehension, cognitive load, and user experience in multilingual meetings. We conducted a within-subjects experiment with 8 bilingual confederates and 47 participants simulating global team meetings with English translations of Greek, Kannada, Mandarin Chinese, and Ukrainian–languages selected for their diversity in grammar, script, and resource availability. Participants experienced four audio conditions: spatial audio with and without background reverberation, and two non-spatial configurations (diotic, monaural). We measured listener comprehension accuracy, workload ratings, satisfaction scores, and qualitative feedback. Spatially-rendered translations doubled comprehension compared to non-spatial audio. Participants reported greater clarity and engagement when spatial cues and voice timbre differentiation were present. We discuss design implications for integrating real-time translation into meeting platforms, advancing inclusive, cross-language communication in telepresence systems.

# Introduction

The story of the Tower of Babel has long symbolized the difficulties of communication across languages. In today's increasingly global and distributed workspaces–where meetings routinely span diverse continents, cultures, and linguistic backgrounds–the echoes of Babel are unmistakable. Multilingual teams are central to innovation, and have enabled landmark scientific and engineering achievements such as the *Human Genome Project* [54], the *International Space Station* [39, 25], and the *CERN Large Hadron Collider* [5]. Nevertheless, language barriers remain a perpetual challenge to seamless collaboration, often requiring live interpreters or machine-mediated conversation captioning [24, 58, 38, 12, 56]. Communication in such settings typically defaults to a single dominant language (a "*lingua franca*") [43, 53], often requiring some participants to operate in a non-native language [41, 17]. Even among proficient second-language speakers, comprehension gaps persist [55, 4], limiting the inclusivity and effectiveness of multilingual meetings. Automated real-time speech translation systems, as found in some commercial voice communication solutions [9, 1], offer a promising bridge by delivering a translated output in the listener's preferred language. However, simply translating the words is not sufficient to create a natural, engaging conversation. Improper integration of translated audio neglecting perceptual design can disrupt the conversational flow, obscure speaker identity, or erode the sense of presence in the meeting [16, 38, 10, 46]. For instance, translated speech might sound like a disembodied voice disconnected from any visible speaker, or it may fully replace the original voice, making it difficult to track who is speaking [35]. These shortcomings raise pressing design questions for delivering translated speech in real time without ruining the user experience.

Meanwhile, the rise of hybrid work has made virtual meetings a default mode of collaboration [46]. Video communication platforms have become the connective tissue of global organizations. Telepresence technologies now aim to recreate the intimacy of in-person interaction, enabling remote participants to feel co-present [33]. Yet, the perceptual design of translated speech in these environments remains underexplored. Prior work has shown that users adapt to machine-mediated conversation but are sensitive to delays, recognition errors, and ambiguous turn-taking and speaker attribution [22, 34]. More recent systems [6] have explored preserving spatial and identity cues in translated speech, demonstrating spatialized audio can improve intelligibility and speaker differentiation, even in noisy environments. In parallel, research on audio augmentation has demonstrated that well-designed auditory cues can convey contextual information effectively without overwhelming the listener [37, 47, 49]. Spatial audio has emerged as a powerful medium for enhancing multi-speaker communication. By rendering sound as if it originates from specific locations in 3D space, spatial audio improves speaker localization, reduces cognitive load, and increases engagement, attention, and comprehension [2, 14, 13]. Prior work has demonstrated the benefits of spatial audio in distributed communication and AR/VR settings [14, 59, 33, 7, 57]. In teleconferencing contexts, spatial audio has been shown to support voice-identification, turn-taking, and increased engagement [40, 27, 28, 30, 46]. However, its application to multilingual meetings–where listeners rely on translated speech to follow speakers they cannot understand–remains largely unexamined.

Building on these insights, we apply spatial audio to a specific and underexplored use case: multilingual videoconferencing with real-time speech translation. We focus on a common scenario: a listener joins a virtual meeting where speakers converse in a language the listener does not understand, and relies on live translation to follow the discussion. In such settings, the perceptual design of translated audio–its spatial placement, voice characteristics, and interaction with visual cues–can significantly influence the listener's ability to understand and engage. Our work bridges this gap by empirically evaluating how spatial audio rendering of live-translated speech affects comprehension, cognitive load, and user experience in multilingual virtual meetings.

We conducted a controlled within-subjects study with 8 bilingual confederates (speak-ers) and 47 participants (listeners), simulating global meetings with Wizard-of-Oz (WoZ) live English translations of conversations in Greek, Kannada, Mandarin Chinese, and Ukrainian. The languages were selected for their diversity in grammar, script, and resource availability. Participants experienced four audio conditions: *spatial* audio for translated speech (aligned with the speaker's on-screen location) with optional reverberation applied to the native speech (to increase perceived speaker distance), and two *non-spatial* configurations (monaural one-ear and monaural two-ears, referred to as *diotic*). We measured comprehension accuracy, NASA-TLX workload ratings, and Likert-scale satisfaction scores, complemented by a thematic analysis from our qualitative feedback.

**Research Questions**   We address two primary research questions:

**RQ1**: *Does spatial audio improve the effectiveness of real-time translation for the listener in terms of comprehension, cognitive load, and satisfaction?*

**RQ2**: *Do spatial cues help the listener better track the conversation dynamics, such as identifying speakers and following turn-taking?*

**Hypotheses**   Based on prior work and our design intuition, we formulated the following hypotheses:

**H1**: **(Comprehension)** Spatial audio will improve comprehension of translated speech compared to non-spatial audio.

**H2**: **(Cognitive Load)** Spatial audio will reduce perceived cognitive load, particularly mental demand and effort.

**H3**: **(Satisfaction)** Spatial audio will increase user satisfaction, including ease of understanding, clarity of speaker distinction, and sense of immersion.

**H4**: **(Speaker Attribution)** Spatial audio will improve the listener's ability to identify who is speaking and map translated speech to the correct speaker.

**Contributions**   This study makes the following contributions:

(i) ***Empirical evidence*** that spatial audio significantly improves comprehension and reduces listening effort in live translated meetings;

(ii) ***Insights*** into how spatial cues and voice timbre differentiation affect user satisfaction and speaker attribution;

(iii) ***Design guideline for integration*** of real-time translation into virtual meetings, emphasizing perceptual coherence.

Together, these contributions advance best practices for inclusive, cross-language communication in telepresence systems and highlight the importance of perceptual design in real-time translation interfaces.

## Related Work

### Real-Time Translation in Conversation

Early work on machine-mediated conversation revealed both the promise and pitfalls of real-time translation. In a formative study by Hara and Iqbal [22], participants engaged in bilingual conversations using a video conferencing system that translated speech on-the-fly.

Users adapted their speaking style (e.g., slowing down, simplifying vocabulary) to accommodate the system. However, persistent recognition and translation errors led to frustration, and turn-taking suffered due to the latency and lack of clear conversational cues. Another study [41] found that audio-only translation yielded better comprehension than audio-plus-video, suggesting that visual cues may not always aid understanding and instead result in a "dubbed movie" effect [31], unless perceptually aligned.

A more recent system has explored preserving spatial and identity cues in translated speech; Chen et al. [6] introduced a wearable translator that rendered translated voices in the direction of the original speaker using binaural audio. Their evaluations showed that spatialized translation improved intelligibility and user experience in realistic noisy environments. They also found that spatial cues aided the preservation of voice identity in translated speech, as reflected by subjective similarity ratings.

These efforts motivate our investigation into how spatial audio can be integrated into conventional video meeting contexts, where design considerations include whether to show the speaker's video, preserve their voice identity, and balance original versus translated channels.

**Audio Augmentation for Contextual Information**

Beyond translation, audio has been used to surface contextual information in ways that support peripheral awareness without overwhelming the listener. Systems such as *Audio Aura* [37] and *Nomadic Radio* [47] pioneered ambient audio cues to deliver background information without interrupting the primary task. These systems used layered soundscapes and voice notifications to convey context such as presence, activity, or message urgency. Later studies found that voice familiarity influenced interruption levels and response times [3, 37]–users reacted fastest to their own voice but found it most disruptive, while unfamiliar voices were less intrusive.

More recent work has extended these ideas to remote collaboration: NAVC [49] used musical audio cues to convey non-verbal feedback (e.g., nodding, smiling) to visually impaired users in video calls. Participants appreciated the added awareness but emphasized the need for accuracy and minimal distraction. Similarly, research on targeted navigation [51, 32], sonified experiences [36, 57], and workspace awareness studies in AR [8] found that audio cues were helpful when paired with visual indicators, increasing engagement and reducing confusion. Auptimize [7] introduced a system for optimizing spatial placement of audio cues to reduce misidentification in AR/VR environments. By leveraging the *ventriloquist effect*, Auptimize improved localization and reduced confusion when multiple sounds occurred simultaneously.

**Spatial Audio in Teleconferencing**

Spatial audio has long been recognized for its benefits in multi-speaker communication. Earlier studies on spatial audio teleconferencing [26, 28] showed that spatial separation helps listeners localize speakers and comprehend overlapping speech better than monaural mixes. More recent work on hybrid teleconferencing [27] found that spatial rendering improves intelligibility and stream identification for both in-room and remote participants, though preferences for spatial width vary by context. Mobile conferencing prototypes [45] combining spatial audio with haptics and gestures were rated as more engaging than traditional setups. In video meetings, spatial audio enhances perceptions of turn-taking and social presence [40]. Participants reported that conversations felt more natural and interactive when voices were spatially rendered.

Psycho-physical studies [14, 2] also show that spatial alignment between audio and visual cues improve selective attention and comprehension. Systems like *Vocal Village* [30] and *Project Starline* [33] demonstrate that spatial audio can reduce listening effort, and increase conversational fluency and presence in virtual meetings. Despite these advances, the application of spatial audio to *multilingual* scenarios remains largely unexplored.

## Materials and Methods

We conducted a controlled within-subjects experiment to evaluate different audio rendering strategies for real-time translation in virtual meetings. The study mimicked a common multilingual scenario: a participant listens to a conversation between two people communicating in an unfamiliar language and relies on an English translation to follow the discussion. To isolate the effects of audio presentation, we manipulated the delivery of translated speech across four conditions and measured outcomes within subjects.

The study was implemented in two phases. **Phase I** involved the recruitment of bilingual confederates who served as speakers in scripted conversations, as well as the development of the audiovisual stimuli–including recording, audio processing, and video stitching. **Phase II** focused on the deployment of the user study via a custom-built web platform, where recruited participants engaged with the pre-recorded conversations and completed comprehension, satisfaction, and workload measures under randomized condition orders. We describe each phase in detail next.

### Phase I: Confederates

**Language Selection**

To ensure that the translated speech was genuinely necessary for comprehension, we selected source languages that are linguistically distinct from English across multiple dimensions. Our selection criteria included language family, syntactic structure (word order), and writing system. *English* (Germanic family, Subject-Verb-Object (SVO) structure, Latin alphabet), the target language for translation, was contrasted with *Greek* (Hellenic family, SVO structure, Greek alphabet), *Kannada* (Dravidian family, Subject-Object-Verb (SOV) structure, Abugida script), *Mandarin Chinese* (Sinitic family, SVO structure, logographic script), and *Ukrainian* (Slavic family, flexible SVO structure, Cyrillic alphabet). These languages were selected for both typological diversity and practical relevance, as Greek, Kannada, and Ukrainian are considered low-resource languages in the context of machine translation [50, 42, 21]. Additionally, Slavic languages such as Ukrainian are highly inflected [10, 11], adding further complexity to translation since the relationship between words is inferred from inflections rather than order. Word order is also a critical factor which directly impacts translation latency in real-time systems. Because Kannada is verb-final (SOV), translation into English (SVO) tends to incur additional lag as interpreters wait for the sentence-final verb [52, 20, 11]. Takahiro Ono and Matsubara [52] quantified this delay, and we sought to simulate similar timing effects in our WoZ setup by modeling longer English offsets for Kannada (SOV) than for Greek, Mandarin Chinese, and Ukrainian (SVO).

**Goal-Based Scenario Design for Scripts**

We designed eight scripted workplace scenarios mirroring common virtual meetings. Instead of full Goal-Based Scenarios (GBS), which emphasize active performance towards a goal [48, 18], we adapted the GBS principle of explicit goal alignment into a goal-referenced assessment blueprint for passive listening. This allowed us to structure each dialogue around clear comprehension targets while acknowledging the *non-interactive* nature of our study. Following established listening assessment practice [4, 19], each ≈1-minute dialogue was designed to elicit evidence for four comprehension goals:

(i) **Temporal Detail** (*when*): recall of time-based facts to assess sequencing, essential for tracking event timelines;

(ii) **Action or Theme** (*what*): grasp of the main topic or purpose, supporting deeper conceptual understanding;

(iii) **Role or Identity** (*who by name*): recognition of participants and their roles, crucial for tracking responsibilities and relationships;

(iv) **Dialogue Attribution** (*who by spatial location*): mapping utterances to the left or right speaker on screen, reinforcing attention to conversational flow and spatial placement in the call.

These goal types balance *surface recall* (e.g., facts, names) with *inferential attribution* (e.g., speaker tracking, role identification), ensuring that comprehension was tested at multiple depths.

In workplace settings, comprehension requires not only following content but also tracking *who* is speaking, *what* they refer to, and *when* effects occur. By mixing surface recall with inferential reasoning, the pattern maintains a consistent difficulty level while remaining accessible to a broad range of participants. It avoids overly technical or abstract questions, making it suitable for general comprehension testing and user experience evaluations. Each dialogue had clear turn-taking and no speech or spatial overlap, i.e., each confederate is in a separate call window (Figure 1.B), resembling realistic workplace exchanges.

To cover a range of common workplace dynamics, we designed dialogues capturing both formal and informal meeting interactions, from high-stakes decision making to everyday coordination, spanning four clusters of scenarios: *onboarding and hosting* (e.g., hiring discussions and welcoming a visitor), *team bonding and informal planning* (e.g., organizing a team-building event and planning a lunch outing), *project coordination* (e.g., giving project updates or preparing a team presentation), and *task management* (e.g., clarifying instructions and addressing delayed tasks). The full scripts, their translations, and multiple-choice questions for each scenario are provided in **Supplemental Material 1**.

**Confederate Recruitment and Wizard-of-Oz Live Translation**

We recruited eight bilingual speakers as confederates, balancing pairs across gender and language groups (Figure 1.B); we ensured to have two *same-gender* pairs (male-male native Ukrainian speakers and female-female native Greek speakers) and two *mixed-gender* (male-female native Kannada and Mandarin Chinese speakers). We recorded four scripts in each of the two gender-groups. Before recording, the confederates were asked to review the translations in their native languages to ensure naturalness and semantic fidelity, and English translations were phrased to sound fluent and idiom-neutral so the study isolates audio rendering rather than translation quality. We used a WoZ approach to simulate a high-quality, stable real-time translator while retaining experimental control over latency, voice identity, and translation accuracy. WoZ is a well-established technique for evaluating interaction concepts before full automation, particularly for speech and language interfaces where live system performance can confound user experience [29, 44]. Besides avoiding the introduction of unpredictable errors or timing variability by an automated machine translation (MT) system, we also made sure to include low-resource languages, some of which are not supported by current MT text-to-speech synthesis.

Sessions were recorded in an acoustically isolated booth (Figure 1.A) using two oriented Sony Alpha 6400 cameras with Sony 16-50mm Power Zoom lens (one per confederate) and Electro-Voice RE-20 microphones. We first recorded English translations (audio-only) for each line, then recorded native-language (audio+video). Confederates rehearsed to ensure natural turn-taking and realistic pacing before final takes. We stitched tracks into multi-channel files (see Figure 1.C), and introduced a scripted delay on English to simulate real-time translation latency while keeping voice identical to the original speaker's timbre, thus, preserving speaker identity across conditions. We varied the translation delay by utterance and language to reflect known word-order effects on simultaneous translation.

**Audio Processing, Spatialization, and Study Conditions**

Each audio channel was rendered using a custom script to apply spatialization effects. All audio channels underwent standard speech enhancement processing to remove room effects and extraneous or unintended noise. Native speech was attenuated by –18 dB relative to the translated speech to ensure the English translation remained foregrounded, and spatialized with a pair of generic Head-Related Transfer Functions (HRTFs) to position sound to the left or right of the listener's head, corresponding to the on-screen position of the active speaker. For the translated speech, we implemented four audio conditions (Figure 2):

 (i) ***Diotic*** (*two-ears*): translated English was played identically in both ears, with no directional cue. This baseline reflects how most commercial real-time translators currently deliver audio.

 (ii) ***Monaural*** (*one-ear*): translated speech was delivered to only one ear (left or right, assigned randomly), while the other ear remained available for residual original speech. We applied HRTFs at azimuth ±90° to avoid discomfort of hard panning. This condition reflects interpreter practices in some professional contexts [15], where users listen to an interpreter through one ear while keeping the other ear open for ambient sound.

 (iii) ***Spatial*** (*two-ears, no reverb*): translated English speech was rendered with spatial cues aligned to the video position of the speaker, without added reverberation.

 (iv) ***Spatial+Reverb*** (*two-ears, with reverb*): same as **(iii)**, but with subtle room reverberation added to the native speech channel to simulate increased perceived distance of the speaker.

## Phase II: Participants

**Study Design**

We employed a within-subjects design with one independent variable: **audio spatialization** condition for translated speech. This factor had four levels: *spatial*, *spatial+reverb*, *diotic*, and *monaural*. Additionally, we extend our analysis to observe differences between audio spatialization modes: ***spatial*** (which includes *spatial* and *spatial+reverb* spatialization modes), and ***non-spatial*** (which includes *diotic* and *monaural* cases). Each of the 16 scripted dialogues (4 languages × 4 scenarios) was rendered under all four audio conditions, resulting in 64 WoZ recordings. To make sure each participant experienced both, same and mixed-gender pairs of confederates and each audio condition, we assigned each participant to one of the four balanced listening groups (*Greek-Kannada*, *Greek-Mandarin Chinese*, *Ukrainian-Kannada*, and *Ukrainian-Mandarin Chinese*). Each group experienced eight dialogues (two pairs × four audio conditions) counterbalanced for order in language and audio condition. To prevent learning effects, each condition featured a different dialogue script.

Before starting, participants calibrated their headphones to confirm left/right alignment and volume comfort. They were instructed that they would hear short, pre-recorded conversations in a language they do not understand, accompanied by live English translation, and that they should pay close attention because they would answer comprehension questions afterward. Following each recording, participants completed a short multiple-choice comprehension quiz and a brief workload and satisfaction survey; after each language session, they provided open-ended feedback in a short debrief. Prior to the study, participants were never told about the audio conditions or content of the recordings. All study procedures were reviewed and approved by the authors' institutional ethics review board and adhered to established ethical guidelines for human-subject research.

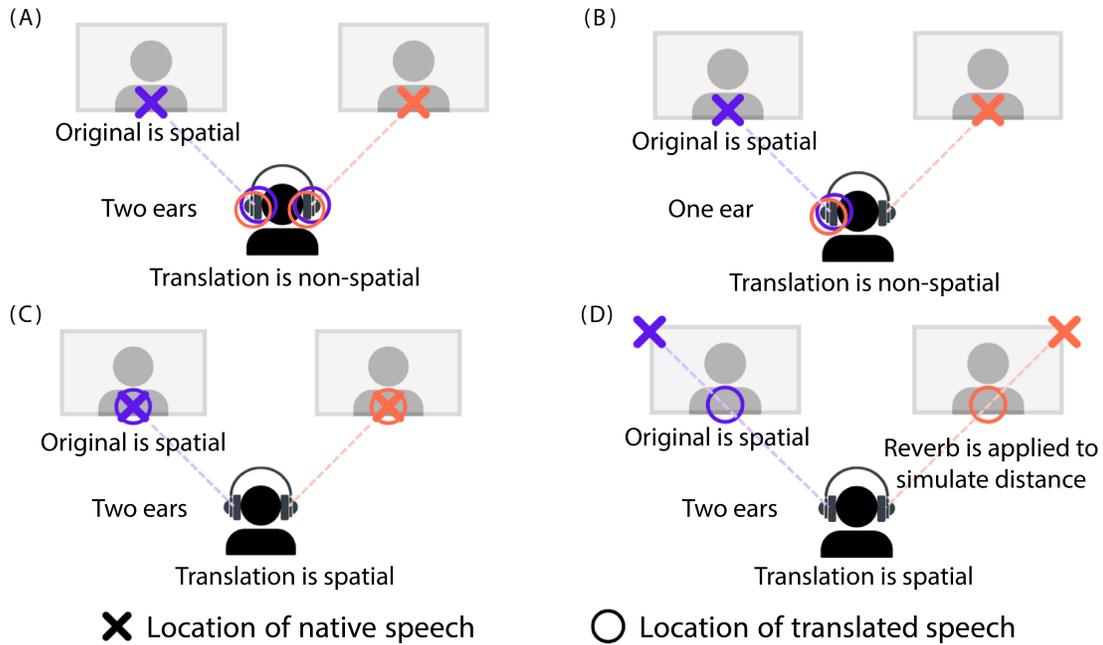

**Figure 2.** Illustration of the four translation audio configurations tested: (A) *diotic*; (B) *monaural*; (C) *spatial+no reverb*; and (D) *spatial+reverb*. Native speech locations are marked with crosses, translated speech with open circles.

**Measures**

We collected both quantitative and qualitative measures in response to the survey questions (see list of satisfaction, workload, and recap questions in **Supplemental Material 2**):

  (i) **Comprehension Accuracy**: for each condition, we calculated the percentage (%) of correctly answered quiz multiple-choice questions. Accuracy was also analyzed by question type. This provided an objective measure of understanding.

 (ii) **Subjective Satisfaction Ratings**: after each condition, participants rated their experience on a Likert scale, focusing on ease of understanding, clarity of voices, and satisfaction with the overall experience.

(iii) **Cognitive Load Questionnaire**: we used the NASA Task Load Index (NASA-TLX) [23], which is a standardized framework to qualitatively assess the perceived load across tasks.

(iv) **Qualitative Feedback**: after each language block, participants provided open-ended reflections in a short debrief, allowing us to capture nuanced preferences, frustrations, and comparisons across conditions.

**Sample Size Determination**

We conducted a Monte Carlo power simulation to determine the required sample size. The logistic regression model used in the simulation excluded random effects to simplify the structure and avoid boundary singular fit issues. For each simulated dataset, we compared a full model including the fixed effect (audio spatialization condition) against a null model (intercept only), using a likelihood ratio test. The resulting *p*-value was used to determine whether the condition effect was statistically significant at the $\alpha = 0.05$ level. This binary outcome was repeated across 500 simulations per sample size to estimate statistical power. The results revealed a clear monotonic increase in power with larger sample sizes (Figure 3) and based on the result, with >45 participants, power exceeded >0.90.

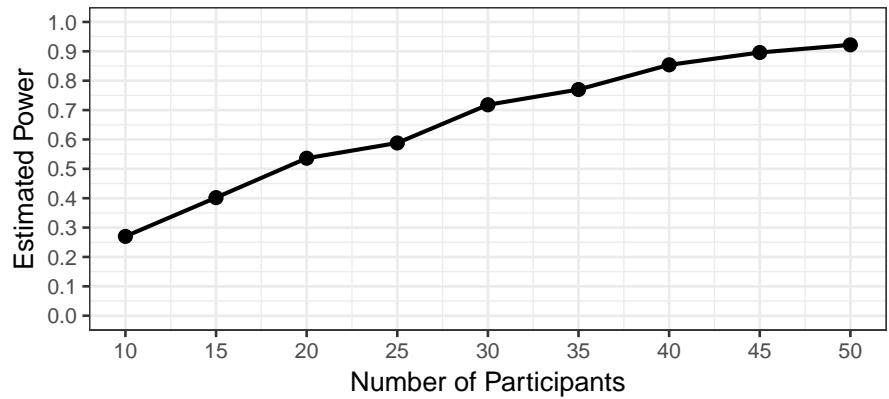

**Figure 3.** Estimated power by sample size using Monte Carlo simulation (500 simulations per sample size).

**Participants**

We recruited 47 participants (24 males, 23 females), all employees at Microsoft Corporation with diverse roles and within the 18-55 age range. The key eligibility criteria were to be fluent in English, and have no to little knowledge of either Greek or Ukrainian, and of Kannada or Mandarin Chinese. Language proficiency was screened using a short online self-assessment questionnaire based on the Interagency Language Roundtable (ILR) scale. Participants rated listening proficiency in *English*, *Hellenic languages* (variants of Greek), *Slavic languages* (Polish, Russian, Ukrainian, etc.), *Dravidian languages* (Kannada, Tamil, Telugu, etc.), and *Sinnitic languages* (Mandarin Chinese, Cantonese, etc.) on a scale from 0 (*No proficiency*) to 5 (*Native Proficiency*). Those who indicated more than 1 (*Elementary Proficiency*) in any source language were excluded. Besides filtering, this onboarding survey allowed us to pair each participant with the most suitable listening group: each one was paired with one same-gender language condition (Greek or Ukrainian) and one mixed-gender language condition (Mandarin Chinese or Kannada) depending on which languages they did not understand. In case of a tie (when they did not know any of the native languages), group assignment was randomized, with balancing to avoid demographic skew. We ended up having 25 Greek, 25 Kannada, 22 Mandarin Chinese, and 21 Ukrainian listeners.

Study sessions were conducted remotely. Participants used a standard laptop running the survey website and wore stereo headphones in a quiet environment. The display showed a simulated video call with the two confederates in their respective video windows. The visual alignment ensured that spatial audio cues (left vs. right ear) corresponded to the perceived speaker on screen. All participants gave informed consent and were compensated with a $25 gift card for about 30 minutes of their time.

# Results

We analyzed comprehension, satisaction, and workload using mixed-effects logistic regression, and performed thematic analysis on the qualitative feedback. To make results interpretable, we report estimated marginal means (EMM) adjusted for random effects, along with pairwise comparisons between conditions.

## Model Structure and Random Effects

Because each participant experienced multiple conditions, responses were not independent, thus, we opted for mixed-effects models to account for this by including both fixed and random effects. *Audio condition* was treated as a fixed effect, while *session order* was modeled as a random intercept to capture variability due to presentation sequence (Figure 4). This approach reduced bias from learning or fatigue across sessions, providing more robust estimates of the true effect of spatial audio.

To select the most appropriate model, we compared five mixed-effects logistic regression models with varying random-effect structures. These included models with random intercepts for *participant ID*, *session order*, both, and random slopes for *session order* nested within *participant ID*. Model fit was evaluated using Akaike Information Criterion (AIC), Bayesian Information Criterion (BIC), and log-likelihood values. The model *Correct ~ AudioType + (1 | SessionOrder)* yielded the lowest AIC and log-likelihood, suggesting it best accounted for variability in presentation order without overfitting. Based on this comparison, we selected that model pattern for reporting the results.

The random intercept captured variability in performance based on the sequence in which participants encountered the conditions. A clear trend emerged (Figure 4): the first session had a negative intercept (< 0), indicating lower baseline performance, while all the others had positive intercepts. Among these, the third session showed the highest improvement by the second one, while the last session, despite being above zero, performed worse than both second and third. This pattern may reflect cognitive fatigue or reduced engagement by the final session, suggesting that presentation order can subtly impact task performance even when not directly manipulated. Including the listening *session order* as a random effect allowed us to control for this variability and isolate the true effect of spatial audio.

## Effects of Spatial Audio

### Comprehension Accuracy

We fitted a mixed-effects logistic regression *Correct ~ AudioType + (1 | SessionOrder)*. Relative to *diotic*, *spatial* audio showed the strongest improvement, with a 113.4% increase in odds of a correct response. This corresponds to an estimated probability of correct response of 75.8% versus 59.5% for the *diotic* baseline. The *spatial+reverb* condition also yielded a statistically significant improvement, with a 48.7% increase in odds (estimated probability 68.6%). In contrast, the *monaural* condition did not differ significantly from the baseline, with a −5.5% decrease in odds, indicating a slight decrease in performance (estimated probability 58.1%). When aggregated into modes, spatial conditions (*spatial* with and without reverb) had an 81.8% increase in odds of a correct response compared to the non-spatial baseline (*diotic* and *monaural*), which results in a 22.76% relative increase in predicted correctness.

To evaluate how different audio conditions influenced comprehension across various question types, we conducted EMM analysis using *Correct ~ AudioType * Question + (1 | SessionOrder)* for modeling (Figure 5). The results, where EMM are presented on the logit scale,

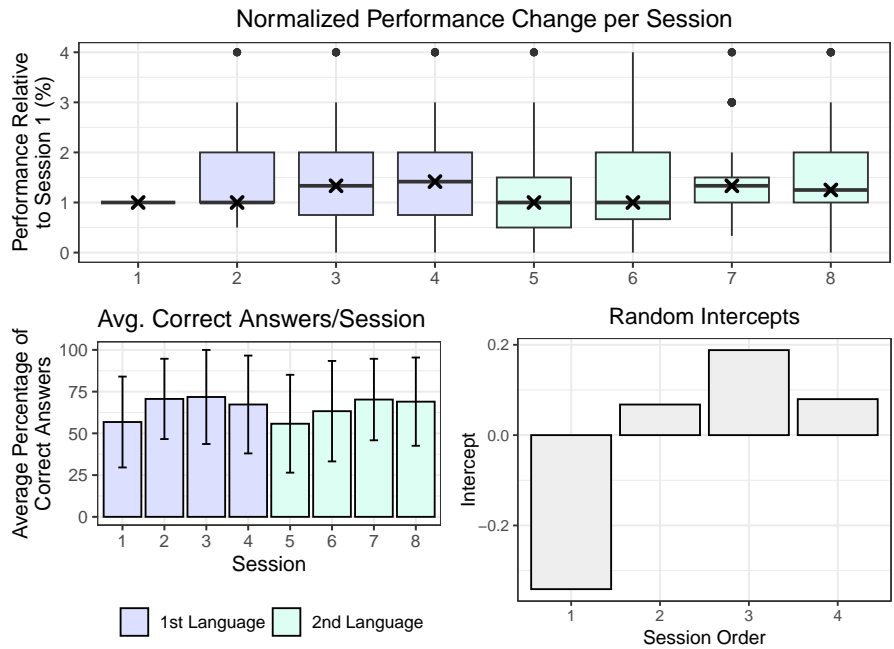

**Figure 4.** Effect of *session order* on comprehension accuracy. In each language block, performance was lowest on the first recording (adaptation), improved in the middle recordings (learnability), and declined again in the final recording (fatigue). This repeating pattern highlights the importance of modeling *session order* as a random effect.

revealed distinct patterns across question categories: for *Temporal Detail* questions, the *spatial* condition yielded the highest predicted probability of correct responses (EMM = 1.09, SE = 0.29 log-odds; corresponding to $\hat{p}$ = 84.0%), followed by *spatial+reverb* (EMM = 0.86, SE = 0.28, $\hat{p}$ = 70.3%), *diotic* (EMM = 0.47, SE = 0.26, $\hat{p}$ = 61.5%), and monaural (EMM = 0.24, SE = 0.26, $\hat{p}$ = 56.0%). While none of the contrasts were statistically significant, the *spatial* condition showed a consistent upward trend, particularly when compared to *monaural* (*monaural - spatial* Δlog-odds = −0.85, SE = 0.36, $z$ = −2.375, $p$ = 0.0819), indicating a potential benefit for time-based comprehension. When grouping in audio condition modes (*spatial* versus *non-spatial*) we have that *spatial* (EMM = 0.97, SE = 0.21, $\hat{p}$ = 72.6%) is notably higher than non-spatial (EMM = 0.35, SE = 0.20, $\hat{p}$ = 58.8%), and their pairwise comparison *non-spatial - spatial* with estimate Δlog-odds = −0.62, SE = 0.25, $z$ = −2.482, $p$ = 0.0131 shows to be statistically significant, suggesting *spatial* mode enhances temporal detail recognition.

For *Action or Theme* questions, performance was highest in the *spatial* condition (EMM = 2.15, SE = 0.39, $\hat{p}$ = 89.5%), followed by *spatial+reverb* (EMM = 1.46, SE = 0.32, $\hat{p}$ = 81.2%), *monaural* (EMM = 1.35, SE = 0.29, $\hat{p}$ = 79.4%), and diotic (EMM = 1.10, SE = 0.29, $\hat{p}$ = 75.0%). Although none of the pairwise comparisons reached statistical significance after adjustment, the spatial condition showed a strong upward trend compared to *diotic* (*diotic-spatial* estimate Δlog-odds = −1.05, SE = 0.46, $z$ = −2.274, $p$ = 0.1042), suggesting a meaningful improvement in conceptual understanding with spatial cues. When grouping in audio condition modes (*spatial* versus *non-spatial*) we have that *spatial* (EMM = 1.77, SE = 0.25, $\hat{p}$ = 85.4%) is notably higher than non-spatial (EMM = 1.22, SE = 0.22, $\hat{p}$ = 77.1%), and their pairwise comparison with estimate Δlog-odds = −0.55, SE = 0.30, $z$ = −1.806, $p$ = 0.0709, thought not statistically significant, suggests that the spatial mode may improve conceptual understanding, but not conclusively.

For *Role or Identity* questions, the *spatial* condition (EMM = 0.52, SE = 0.26, $\hat{p}$ = 62.7%) showed a significant improvement over *diotic* (EMM = −0.42, SE = 0.26, $\hat{p}$ = 39.6%) with a pairwise comparison *diotic - spatial* estimate Δlog-odds = –0.94, SE = 0.34, $z$ = –2.766, $p$ = 0.0290, and *monaural* (EMM = –0.48, SE = 0.26, $\hat{p}$ = 38.3%) with a *monaural - spatial* estimate of Δlog-odds = –1.00, SE = 0.34, $z$ = –2.919, $p$ = 0.0184, suggesting that spatial audio significantly enhances semantic role identification. When grouping in audio condition modes we have that *spatial* mode (EMM = 0.307, SE = 0.20, $\hat{p}$ = 57.6%) is notably higher than *non-spatial* (EMM = 0.31, SE = 0.20, $\hat{p}$ = 38.9%), and their pairwise comparison with estimate Δlog-odds = –0.76, SE = 0.24, $z$ = –3.173, $p$ = 0.0015 shows to be statistically significant, indicating that the spatial mode significantly improves semantic role identification.

In *Dialogue Attribution* questions, the *spatial* condition again yielded the highest estimated mean (EMM = 1.21, SE = 0.53, $\hat{p}$ = 77.1%), followed by *spatial+reverb* (EMM = 1.05, SE = 0.49, $\hat{p}$ = 74.0%), *diotic* (EMM = 0.69, SE = 0.45, $\hat{p}$ = 66.7%), and monaural (EMM = 0.41, SE = 0.45, $\hat{p}$ = 60.1%), but all contrasts were non-significant. When grouping in audio condition modes we have that *spatial* mode (EMM = 1.12, SE = 0.37, $\hat{p}$ = 75.5%) is notably higher than *non-spatial* (EMM = 0.55, SE = 0.32, $\hat{p}$ = 63.5%), and their pairwise comparison with estimate Δlog-odds = –0.57, SE = 0.46, $z$ = –1.227, $p$ = 0.2199 shows no significance, so no strong evidence that spatial mode improves speaker attribution.

**Satisfaction Ratings**

Next, we examined subjective satisfaction across five dimensions of perceived experience: *ease of understanding the conversation* ($S_1$), *clarity of speaker distinction* ($S_2$), *sense of immersion and engagement* ($S_3$), *distraction from the original audio* ($S_4$), and *overall listening experience* ($S_5$). Each model was fitted using either a mixed effects structure with the listening session order as a random intercept ($S_1$, $S_4$, and $S_5$) or a simpler linear model when singularity was detected ($S_2$ and $S_3$), signaling that the variability of *session order* was negligible. Results revealed that spatial audio significantly improved user experience in the first three dimensions. Specifically, participants rated the *spatial* mode as easier to understand ($S_1$: *non-spatial - spatial* Δlog-odds = –0.38, SE = 0.15, $z$ = –2.584, $p$ = 0.0098), clearer in distinguishing speakers ($S_2$: *non-spatial - spatial* = Δlog-odds = –0.543, SE = 0.21, $z$ = –2.578, $p$ = 0.0103), and more immersive ($S_3$: *non-spatial - spatial* Δlog-odds = –0.387, SE = 0.16, $z$ = –2.479, $p$ = 0.0136) compared to *non-spatial* audio. These effects were consistent across models, with spatial audio yielding higher EMM in each case. Among *non-spatial* conditions, we can observe that for $S_3$, *diotic* has a significative preference over *monaural* (*diotic - monaural* Δlog-odds = 0.57, SE = 0.22, $z$ = 2.598, $p$ = 0.0478). In contrast, no significant differences were observed for distraction ($S_4$: *non-spatial - spatial* Δlog-odds = 0.07, SE = 0.18, $z$ = 0.375, $p$ = 0.7075) or overall satisfaction ($S_5$: *non-spatial - spatial* Δlog-odds = –0.22, SE = 0.15, $z$ = –1.491, $p$ = 0.1369), suggesting that while spatialization enhances cognitive and perceptual clarity, it may not strongly influence emotional or holistic satisfaction.

When analyzing the random intercepts for satisfaction measures across sessions, we observed distinct patterns that offer insight into how participants' experiences evolved over time. In $S_1$ the random intercepts showed a slight increase with each subsequent session. This upward trend suggests that participants found the task progressively easier as they became more familiar with the system and its translation behavior. In $S_4$ the random intercepts decreased over time, indicating that participants felt less distracted in later sessions. This trend likely reflects perceptual habituation: as users became accustomed to the voices, accents, and audio mixing characteristics of the speakers, their ability to filter out irrelevant or competing auditory information improved. Finally, $S_5$ showed an increasing trend in random intercepts across sessions. This suggests that participants' holistic evaluation of the system improved over time, possibly due to growing familiarity, increased confidence in interpreting the translated content, and reduced effort in navigating the interface.

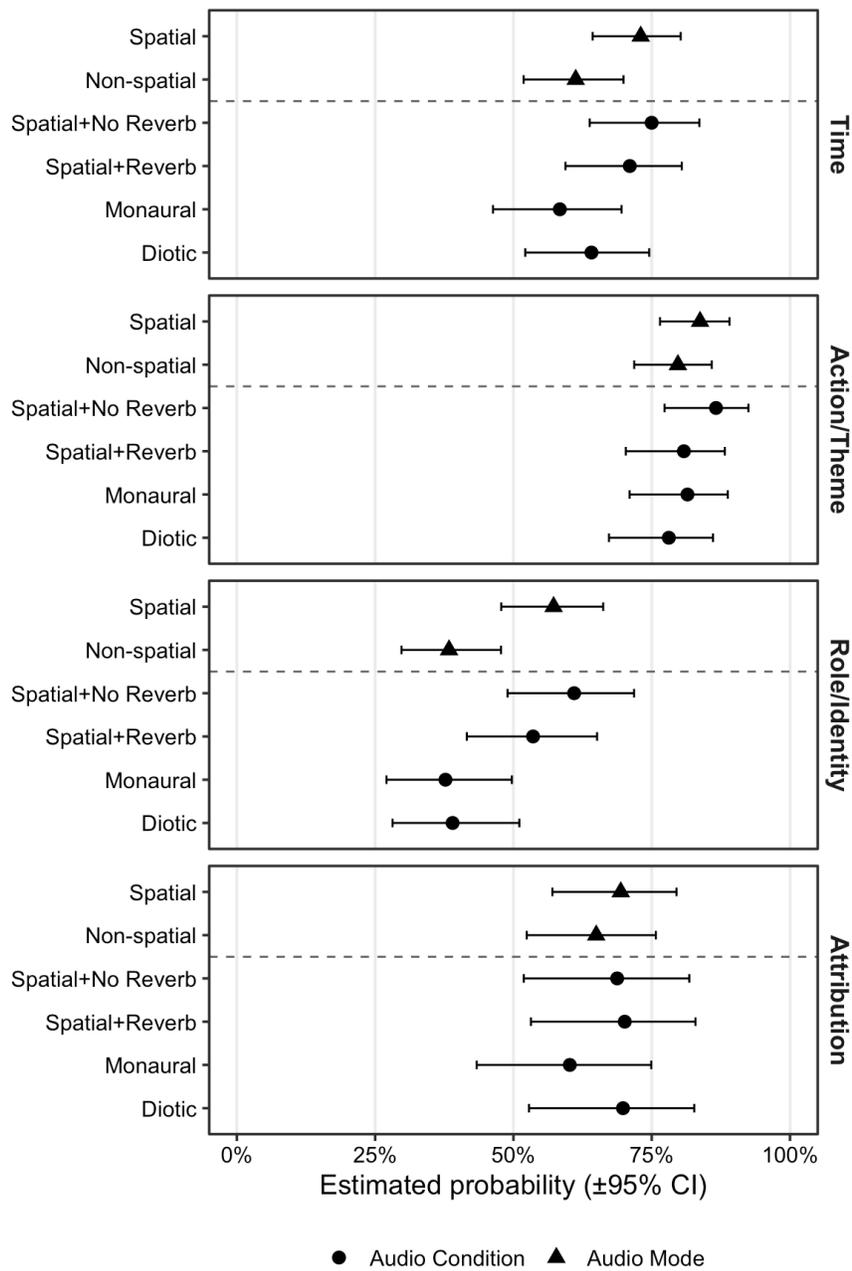

**Figure 5.** Estimated probabilities (±95% CI) of correct comprehension for four information types (*Temporal Detail*, *Action or Theme*, *Role or Identity*, *Dialogue Attribution*) under different translation audio renderings. Circles represent specific audio conditions (*spatial*, *non-spatial*, *spatial+ono reverb*, *spatial+reverb*, *monaural*, *diotic*), while triangles indicate aggregated audio mode effects (*spatial* vs. *non-spatial*). Spatially rendered translations consistently improved comprehension across categories compared to non-spatial conditions.

**Workload**

Spatial audio showed consistent but subtle reductions in perceived workload across our four NASA-TLX dimensions: *mental demand* ($TLX_1$), *temporal demand* ($TLX_2$), *effort* ($TLX_3$), and *performance* ($TLX_4$). Each fitted model compared *spatial* versus *non-spatial* audio modes using either linear regression or mixed effects modeling with *session order* as a random intercept when appropriate. Across $TLX_1$-$TLX_3$, which reflect cognitive and attentional load, spatial audio consistently trended toward lower workload ratings, though none of these differences reached statistical significance ($TLX_1$: *non-spatial - spatial* Δlog-odds = 0.215, SE = 0.14, $t(370)$ = 1.572, $p$ = 0.1168; $TLX_2$: *non-spatial - spatial* Δlog-odds = 0.15, SE = 0.17, $z$ = 0.861, $p$ = 0.3890, and $TLX_3$: *non-spatial - spatial* Δlog-odds = 0.161, SE = 0.14, $t(370)$ = 1.132, $p$ = 0.2583). These results suggest that while spatial audio may reduce perceived mental effort and time pressure, the effect is subtle and may be influenced by individual differences or task familiarity. In contrast, $TLX_4$—measuring participants' confidence in their performance—revealed a statistically significant increase in ratings under spatial audio ($TLX_4$: *non-spatial - spatial* Δlog-odds = –0.364, SE = 0.15, $z$ = –2.555, $p$ = 0.0106). This indicates that users felt more assured in their ability to understand key points when spatial cues were present. This finding highlights the role of spatialization not only in reducing cognitive friction but also in enhancing users' sense of control and success in complex listening tasks. The lack of strong effects in $TLX_1$ and $TLX_3$ may reflect the adaptability of users to both audio conditions, while in $TLX_2$ the trend of random intercepts is decreasing over listening sessions, and in $TLX_4$ it is increasing.

**Associations between Satisfaction and Workload**

To explore associations between variables, we analyzed the angles between PCA biplot vectors [18], which represent the loadings of each factor on the principal components. In this context, the angle between two vectors serves as a geometric proxy for the correlation between the corresponding variables. Specifically, smaller angles (approaching 0°) indicate strong positive correlations, while angles near 90° suggest orthogonality, implying little to no linear relationship. Our results revealed several variable pairs with angles less than 10°, including $S_1$–$S_3$, $TLX_1$–$TLX_3$, and $TLX_1$–$TLX_5$, all of which exhibited cosine values above 0.99. These findings suggest that these variables are highly aligned in the PCA space and likely measure similar underlying constructs or exhibit similar response patterns across participants. Conversely, we identified a distinct set of variable pairs with angles between 85° and 95°, such as $TLX_3$–$S_2$, $TLX_2$–$S_3$, and $TLX_1$–$TLX_4$, which showed cosine values close to zero. These near-orthogonal relationships imply that the variables contribute independently to the principal components and capture distinct dimensions of the data.

## Effects of Voice Timbre

We also examined whether voice pairing influenced comprehension. We fitted a generalized linear mixed model with a binomial link function, modeling correct responses as a function of voice pairing (same-gender vs. mixed-gender) and including an interaction with question type, and the listening *session order* as a random intercept, *Correct ~ VoiceType * Question + (1 | SessionOrder)*. The results revealed a significant interaction for the *Dialogue Attribution* question ($p$ < 0.0001), which asked participants to identify who said what. In this case, same-gender voice pairs (EMM = –0.15, SE = 0.26, $\hat{p}$ = 46.2%) led to significantly lower accuracy compared to mixed-gender (EMM = 1.75, SE = 0.33, $\hat{p}$ = 85.2%), suggesting that vocal similarity may hinder speaker differentiation. For other question types (*Role or Identity*, *Temporal Detail*, and *Action or Theme*) no significant interactions were observed.

### Inductive Thematic Analysis

We coded 376 open-ended responses to our recap questions from 47 participants, where four themes emerged. In what follows, we report prevalence, mention, findings and include selected key quotes:

**Theme 1: Timing and Overlap Break Conversational Flow** (33/47 participants (70.2%), 47 mentions)

Participants frequently described translation lag and overlap between native and translated speech as the primary barriers to following the discussion. Lag made it difficult to align the voice they heard with the person on screen: *"with this offset, I find it incredibly difficult to focus on the specifics of what people are saying, because the facial expressions and body language do not match the audio timing I am hearing"* (P7599), and overlap forced them to split attention across competing streams, often leading to missed details: *"even though I don't understand the original language, I feel distracted to hear two voices of different persons talking at the same time"* (P5732). Additionally, participants reported masking effects due to overlap: *"having the original audio in the background sometimes masked the English translated audio"* (P8866). Several participants reported coping by looking away or closing their eyes to focus on the audio alone: *"at one point it was easier for me to close my eyes and focus on the sound as it was extremely distracting to hear the original audio [...] at the same time that I am trying to pay attention on the translated audio"* (P7257). These accounts concur with our quantitative findings that *session order* influences performance as participants eventually learn some coping mechanisms.

**Theme 2: Audio Configuration Shapes Cognitive Effort** (27/47 participants (57.4%), 47 mentions)

Most participants described the one-ear configurations as mentally taxing, *"unnatural"* (P3524), or even *"jarring"* (P6841) and reported fatigue when concentrating on a single channel–especially with quick speaker changes: *"I found it difficult to focus on just one ear, and when the speakers switched quickly, I sometimes lost track of the conversation because I couldn't always identify who was speaking"* (P5631). However, a minority reported that the monaural mode actually made focusing easier: *"I think it was the easiest. It seemed that the translated audio was coming from a separate location than the original audio, and this made the original audio easier to ignore"* (P1669). Spatial conditions were generally perceived as if the native speech track was attenuated because the translated channel felt dominant and, consequently, focus was improved. In general, participants agreed that lowering the background speech would aid reducing the cognitive workload: *"having to remember key points and constantly trying to filter out the original speech from the interpretation [is] difficult"* (P1123). These accounts align with NASA-TLX trends (lower mental and effort ratings under spatial) and the significant boost in performance with spatial audio.

**Theme 3: Spatial and Timbre Cues Enable Speaker Identification** (27/47 participants (57.4%), 38 mentions)

Listeners repeatedly emphasized the value of left/right spatial mapping: *"I liked it better when I could hear one speaker in each era; it made it easier to distinguish who was saying what and correlate speaker with the translated voice"* (P7267), and distinct voice timbres, e.g., gender-differentiated voices: *"I was able to assess who was speaking based on the differences of a male and female voice, making the conversation easier to follow"* (P2136) for tracking *who said what* (P7832). Spatial separation made turn-taking easier to follow and reduced confusion during rapid switches; conversely, similar-sounding voices or same-gender pairings impeded attribution: *"the fact that the voices were very similar made it super hard for me to distinguish between the speakers"* (P9090). This maps onto our quantitative result that spatial audio significantly improved *Role* identification, and it contextualizes the observed effect of voice timbre with *Dialogue Attribution* accuracy.

**Theme 4: Two-ear and Spatial Configurations Aid Comprehension** (15/47 participants (31.9%), 20 mentions)

A subset explicitly attributed better understanding to hearing with both ears or to spatial alignment of the translated voice with the on-screen speaker: *"listening with both ears helped me understand better. It just felt easier to understand and try to connect the voices with the people"* (P1051). Others noted the same benefit indirectly, e.g., *"more natural"* (P310), *"more confident with responses"* (P6268), and *"helpful for understanding"* (P9760). The lower prevalence here is consistent with the tendency for participants to describe mechanisms (such as timing, masking, attribution) rather than label the outcome (comprehension)–which our quantitative analyses nonetheless confirm (higher odds of correct responses under spatial conditions).

# Discussion

Across sessions, spatial audio improved comprehension relative to non-spatial delivery, with the largest and statistically significant gains for semantic *Role* identification (*who said what*). Satisfaction mirrored this pattern–participants rated spatial modes as easier to understand, clearer for distinguishing speakers, and more immersive–and self-rated performance (NASA-TLX) was significantly higher under spatial conditions. One-ear configurations were frequently judged unnatural and mentally taxing, aligning with the absence of binaural advantages and increased effort to maintain a single-channel focus while tracking turn-taking. The qualitative results explain the mechanism behind these effects: the most prevalent concern was temporal misalignment (70.2% participants), followed by cognitive effort (57.4%), and speaker differentiation (57.4%). A smaller subset explicitly named comprehension (31.9%), but many implied it through mechanisms (e.g., lag, masking, attribution). Together, these patterns suggest that the benefit of spatialization operates through ***stream segregation***, i.e., reducing informational masking, and ***audiovisual coherence***, that is, linking sound to the right face, but these can be blunted by timing errors and overlap that force listeners to split attention.

## Implications for Design

### Spatial Mapping as a Default

Spatializing the translated voice to the speaker's on screen-side improves both understanding and speaker attribution in multilingual virtual meetings. In practical terms, one should keep stable left/right panning per talker within a session, and avoid monaural one-ear as a default–most participants found it confusing or fatiguing when listening includes an exchange between multiple participants.

### Balancing Background Speech Level

Participants repeatedly described masking from the native speech track. A robust default is to duck the original speech while translation is active. While we used –18 dB attenuation, some participants still reported it to be loud and a few had a perceptual sensation as if the spatial condition increased the level gap between the original and translated tracks. The recommendation is to provide a single "original ↔ translated" balance slider so users can adapt to context and hearing preferences.

### Identity Matters: Differentiate Timbre by Speaker

Same-gender and similar-timbre pairings reduced dialogue attribution accuracy. If using MT, the recommendation is to choose contrasting voices and maintain identity binding throughout the meeting by keeping them consistent across turns.

### Controls and Onboarding that Respect Adaptation

Users improved with exposure, as clearly noted by the *session order* effects. Lightweight onboarding would flatten the learning curve and can be done by a left/right test, a quick level calibration, and one-line explanation that translation may arrive slightly after speech. One may add a "focus mode": with spatialized translation, attenuated original, and optional video dimming–many participants voluntarily looked away to focus on audio during lag.

## Limitations and Future Work

Latency (i.e., both lag and overlap) remains the largest pain point and a first-class constraint for improving real-time translation. Our WoZ translation was near-optimal in timing and semantics, but real systems will inevitably introduce automatic speech recognition or MT errors that may interact with spatialization. Another limitation is that our scenarios involved the specific case of two speakers and headphone listening: multi-party meetings where even more voices are involved, loud or reverberant environments, and speakerphones may change the cost-benefit balance. Besides, we used pre-recorded dialogues that excluded live dynamics such as interruptions, laughter, or turn-grabbing, which may exacerbate overlap challenges. Finally, we used generic HRTFs; personalized HRTFs or head-tracking might further improve attribution.

We foresee several promising short-term directions: studying more in depth end-to-end deployments in live meetings including more than two participants, interactive exchanges, and realistic network jitter; exploring adaptive mixing with speaker-change detection to automatically duck competing streams; personalizing the translation/original balance over time and, finally, examining how different voice timbres (gendered, accented, synthetic) affect trust, perceived authority, and inclusion, with attention to equitable defaults.

## Conclusions

Designing translated meetings is not only a language problem—it is a perception and attention problem. Spatial alignment and identity cues reliably improve understanding, but timing discipline, i.e., lag and overlap minimization, is the make-or-break factor. Systems that treat *space*, *time*, and *identity* as co-equal design surfaces will better support inclusive, cross-language collaboration.

## Acknowledgments

We would like to thank all our study participants. Special thanks to Matt McGinley, Deeksha M. Shama, Heng Du, Jana Casals, Viktor Golub, Andrii Iermolaiev, Dimitra Emmanoilidou, and Sebastian Braun.

# Supporting Information

# 1  Scripts

## Hiring Discussion

**English translation**

> **Speaker 1:** Hi Alex, have you had a chance to review the applications for the UX designer role?
> **Speaker 2:** Yes, I went through them yesterday. We've got three strong candidates. I think we should interview all three.
> **Speaker 1:** Great. Did anyone stand out to you?
> **Speaker 2:** I was really impressed by Priya Sharma. Although her background is in software engineering, she has five years of experience in UX design and worked on a similar product in her own startup.
> **Speaker 1:** That's promising. What about availability?
> **Speaker 2:** I've already reached out. She's available for an interview this Thursday at 2 PM.
> **Speaker 1:** Perfect. Let's schedule it. And let's also make sure we include a design task in the interview.
> **Speaker 2:** Agreed. I'll draft something simple but relevant to our product.
> **Speaker 1:** Sounds good. Let's regroup after the interviews to compare notes.

**Comprehension Quiz**

**Temporal Detail (Time-based fact)**   *When is Priya's interview scheduled?*

   A) Wednesday at 2 PM

   B) **Thursday at 2 PM** ✓

   C) Friday at 10 AM

   D) Thursday at 10 AM

**Action or Theme (Conceptual understanding)**   *What will be included in the interview?*

   A) A coding test

   B) A group discussion

   C) **A design task** ✓

   D) A presentation

**Role or Identity (Semantic reference)**   *What is Priya Sharma's background?*

   A) Marketing

   B) UX Design

   C) **Software Engineering** ✓

   D) HR

**Dialogue Attribution (Speaker reference)**  *Who reached out to Priya Sharma about availability?*

- A) Speaker on the left side of the screen
- B) **Speaker on the right side of the screen** ✓
- C) Both
- D) None

**Text in Kannada**

**Speaker 1:** ಹಾಯ್ alex, UX ವಿನ್ಯಾಸಕ ಹುದ್ದೆಗೆ ಅರ್ಜಿಗಳನ್ನು ಪರಿಶೀಲಿಸಲು ನಿಮಗೆ ಅವಕಾಶ ಸಿಕ್ಕಿದೆಯೇ?
**Speaker 2:** ಹೌದು, ನಾನು ನಿನ್ನೆ ಅವುಗಳನ್ನು ಪರಿಶೀಲಿಸಿದೆ. ನಮ್ಮಲ್ಲಿ ಮೂವರು ಪ್ರಬಲ ಅಭ್ಯರ್ಥಿಗಳಿದ್ದರೆ . ನಾವು ಮೂವರನ್ನು ಸಂದರ್ಶಿಸಬೇಕು ಎಂದು ನನಗೆ ಅನಿಸುತ್ತದೆ.
**Speaker 1:** ಅದ್ಭುತ. ಯಾರಾದರೂ ನಿಮಗೆ ಎದ್ದು ಕಾಣಿದರೆ?
**Speaker 2:** ಪ್ರಿಯ ಶರ್ಮ ನಿಂದ ನಾನು ನಿಜವಾಗಿಯೂ ಪ್ರಭಾವಿತನಾದೆ. ಅವಳ ಹಿನ್ನೆಲೆ ಸಾಫ್ಟ್‌ವೇರ್ ಎಂಜಿನಿಯರಿಂಗ್‌ಲ್ಲಿದ್ದರೂ , ಅವಳು UX ವಿನ್ಯಾಸದಲ್ಲಿ ಐದು ವರ್ಷಗಳ ಅನುಭವವನ್ನು ಹೊಂದಿದ್ದಾಳೆ ಮತ್ತು ತಮ್ಮದೇ ಆದ ಸ್ಟಾರ್ಟಪ್‌ಪ್ಪಲ್ಲಿ ಇದೆ ರೀತಿಯ ಉತ್ಪನ್ನದ ಮೇಲೆ ಕೆಲಸ ಮಾಡಿದ್ದಾಳೆ
**Speaker 1:** ಅದು ಭರವಸೆ ನೀಡುತ್ತದೆ.ಲಭ್ಯತೆಯ ಬಗ್ಗೆ ಏನು?
**Speaker 2:** ನಾನು ಆಗಲೇ ಸಂಪರ್ಕಿಸಿದ್ದೇನೆ. ಅವಳು ಈ ಗುರುವಾರ ಮಧ್ಯಾಹ್ನ 2 ಗಂಟೆಗೆ ಸಂದರ್ಶನಕ್ಕೆ ಲಭ್ಯವಿದ್ದಾಳೆ
**Speaker 1:** ಒಳ್ಳೆಯದ್ಯೈ. ಅದನ್ನು ನಿಗದಿ ಮಾಡೋಣ. ಮತ್ತು ಸಂದರ್ಶನದಲ್ಲಿ ಒಂದು ವಿನ್ಯಾಸ ಕಾರ್ಯವನ್ನು ಖಂಡಿತವಾಗಿ ಸೇರಿಸೋಣ
**Speaker 2:** ಒಪ್ಪುತ್ತೇನೆ. ನಾನು ಸರಳವಾದ ಆದರೆ ನಮ್ಮ ಉತ್ಪನ್ನಕ್ಕೆ ಸಂಬಂಧಿಸಿದ ಏನನ್ನಾದರೂ ರಚಿಸುತ್ತೇನೆ
**Speaker 1:** ಚೆನ್ನಾಗಿದೆ. ಟಿಪ್ಪಣಿಗಳನ್ನು ಹೋಲಿಸಲು ಸಂದರ್ಶನಗಳ ನಂತರ ಮರುಸಂಗ್ರಹಿಸೋಣ

**Text in Mandarin Chinese**

**Speaker 1:** 你好，Alex，你有机会看一下 UX 设计师职位的申请吗？
**Speaker 2:** 是的，我昨天看过了。我们有三个很有实力的候选人。我觉得我们应该把他们三个都面试一下。
**Speaker 1:** 太好了。有谁让你印象深刻吗？
**Speaker 2:** Priya Sharma 给我留下了深刻的印象。虽然她的背景是软件工程，但她在用户体验设计方面有五年的经验，并且在自己的初创公司开发过类似的产品。
**Speaker 1:** 听上去不错。她什么时间有空呢？
**Speaker 2:** 我已经联系她了。她可以在本周四下午 2 点接受面试。
**Speaker 1:** 太好了。我们安排一下吧。另外，我们需要确保在面试中加入一项设计任务。
**Speaker 2:** 我同意。我会起草一些简单但与我们产品相关的内容。
**Speaker 1:** 听起来不错。面试结束后我们再碰面交流一下结果

## Planning of a Teambuilding Event

### English translation

**Speaker 1:** Good evening, Mia, any ideas for our next team-building event?
**Speaker 2:** Hello Jerry, I was thinking of a cooking class. It's interactive and fun, and we can do it in-person or virtually.
**Speaker 1:** That sounds great. But we should check for dietary restrictions.
**Speaker 2:** Good point. I know Alex is allergic to shellfish, and Dana is vegan.

**Speaker 1:** Then maybe we should go with a plant-based menu. That way everyone can participate.
**Speaker 2:** Agreed. I'll reach out to a local chef who specializes in vegan cooking.
**Speaker 1:** Awesome. What date are you thinking?
**Speaker 2:** How about the 18th? It's a Friday, and most people are free in the afternoon.
**Speaker 1:** Perfect. Let's send out a quick RSVP form to confirm.

**Comprehension Quiz**

**Temporal Detail (Time-based fact)** *What date is proposed for the event?*

- A) The 15th
- B) **The 18th** ✓
- C) The 20th
- D) The 25th

**Action or Theme (Conceptual understanding)** *What kind of menu will the event have?*

- A) Seafood
- B) **Vegan** ✓
- C) Italian
- D) Gluten-free

**Role or Identity (Semantic reference)** *Who is allergic to shellfish?*

- A) Dana
- B) Mia
- C) **Alex** ✓
- D) Jerry

**Dialogue Attribution (Speaker reference)** *Who suggested a cooking class as a team-building event?*

- A) **Speaker on the left side of the screen** ✓
- B) Speaker on the right side of the screen
- C) Both
- D) None

**Text in Kannada**

**Speaker 1:** ಶುಭ ಸಂಜೆ,ಮಿಯಾ, ನಮ್ಮ ಮುಂದಿನ ತಂಡ ನಿರ್ಮಾಣ ಕಾರ್ಯಕ್ರಮಕ್ಕೆ ಏನಾದರು ಉಪಾಯಗಳಿವೆಯೇ?
**Speaker 2:** ಹಾಲೋ ಜೆರ್ರಿ, ನಾನು ಅಡುಗೆ ತರಗತಿಯ ಬಗ್ಗೆ ಯೋಚಿಸುತ್ತಿದ್ದೆ. ಇದು ಸಂವಾದಾತ್ಮಕ ಮತ್ತು ಮೋಜಿನ ಸಂಗತಿಯಾಗಿದೆ, ಮತ್ತು ನಾವು ಅದನ್ನು ಯ್ಯೆಯುಕ್ತಿಕವಾಗಿ ಅಥವಾ ವಾಸ್ತವಿಕವಾಗಿ ಮಾಡಬಹುದು
**Speaker 1:** ಅದು ಚೆನ್ನಾಗಿದೆ. ಆದರೆ ನಾವು ಆಹಾರದ ನಿರ್ಬಂಧಗಳನ್ನು ಪರಿಶೀಲಿಸಬೇಕು
**Speaker 2:** ಒಳ್ಳೆಯ ವಿಷಯ. Alex ಇಗೆ ಚಿಪ್ಪು ಮೀನು ಗಳಿಂದ ಅಲರ್ಜಿ ಇದೆ ಎಂದು ನನಗೆ ತಿಳಿದಿದೆ ಮತ್ತು Dana vegan ಆಗಿದ್ದಾಳೆ.
**Speaker 1:** ಹಾಗಾದರೆ ನಾವು ಗಿಡ ತಾರಕಾರಿ ಮೂಲ್ಯಗಳ ಆಹಾರುವಿನೊಂದಿಗೆ ಹೋಗಬಹುದು. ಆ ರೀತಿಯಲ್ಲಿ ಎಲ್ಲರು ಭಾಗವಹಿಸಬಹುದು
**Speaker 2:** ಒಪ್ಪುತ್ತೇನೆ. ನಾನು vegan ಅಡುಗೆಯಲ್ಲಿ ಪರಿಣತಿ ಹೊಂದಿರುವ ಸ್ಥಳೀಯ ಬಾಣಸಿಗರನ್ನು ಸಂಪರ್ಕಿಸುತ್ತೇನೆ
**Speaker 1:** ಅದ್ಭುತ. ನೀವು ಯಾವ ದಿನಾಂಕವನ್ನು ಯೋಚಿಸುತ್ತಿದ್ದೀರಿ?
**Speaker 2:** 18 ನೇ ತಾರೀಖಿನ ಬಗ್ಗೆ ಹೇಗೆ? ಇದು ಶುಕ್ರವಾರ, ಮತ್ತು ಹೆಚ್ಚಿನ ಜನರು ಮಧ್ಯಾಹ್ನ ಲಭ್ಯವಿದ್ದಾರೆ.
**Speaker 1:** ಒಳ್ಳೆಯದು. ನಾವು ದೃಢಪಡಿಸಲು ತುರಂತವಾಗಿ RSVP ಅರ್ಜಿಅನ್ನು ಕಳಿಸೋಣ

**Text in Mandarin Chinese**

**Speaker 1:** 晚上好，米娅，下次团队建设活动有什么想法吗？
**Speaker 2:** 你好，杰瑞，我想开一个烹饪课。它互动性强，很有趣，我们可以线下或线上进行。
**Speaker 1:** 听起来不错。但我们应该确认一下忌口情况。
**Speaker 2:** 说得对。我知道亚历克斯对贝类过敏，而达娜是纯素食者。
**Speaker 1:** 那我们或许可以尝试一下素食菜单。这样每个人都可以参加。
**Speaker 2:** 同意。我会联系一位擅长纯素食烹饪的当地厨师。
**Speaker 1:** 太棒了。你打算哪天？
**Speaker 2:** 18 号怎么样？那天是星期五，大多数人下午都有空。
**Speaker 1:** 太好了。我们赶快发送一份确认邮件。

## Hosting a Visitor

### English translation

**Speaker 1:** Hey Sam, did you hear that Professor Lin is visiting next week?
**Speaker 2:** Hi Kim, yes! She's giving a talk on AI ethics on Tuesday at 11 AM.
**Speaker 1:** Do we have her accommodation sorted?
**Speaker 2:** Yep, she's staying at the Redmond Inn. It's close to campus and has a shuttle.
**Speaker 1:** Great. Are we doing a lunch afterward?
**Speaker 2:** Yes, we've booked a table at the Garden Café. It's quiet and has vegetarian options.
**Speaker 1:** Perfect. Who's picking her up from the airport?
**Speaker 2:** I am. Her flight arrives Monday at 6 PM.
**Speaker 1:** Thanks, Sam. Let's make sure she has everything she needs.

### Comprehension Quiz

**Temporal Detail (Time-based fact)**   *When is Professor Lin arriving in town?*

A) **Monday at 6 PM** ✓

B) Tuesday at 11 AM

C) Tuesday at 6 PM

D) Monday at 11 AM

**Action or Theme (Conceptual understanding)**   *Where is Professor Lin staying during her visit?*

A) Garden Café

B) **Redmond Inn** ✓

C) Redmond Hotel

D) Bellevue Suites

**Role or Identity (Semantic reference)**   *Why is Professor Lin visiting?*

A) **To give a talk on AI Ethics** ✓

B) To lead a workshop on Robotics and AI

C) To attend a Machine Learning showcase

D) To meet with the Data Privacy team

**Dialogue Attribution (Speaker reference)**   *Who is picking Professor Lin up from the airport?*

A) Speaker on the left side of the screen

B) **Speaker on the right side of the screen** ✓

C) Both

D) None

**Text in Kannada**

> **Speaker 1:** ಹೇ ಸ್ಯಾಮ್, ಪಪ್ರೊಫೆಸರ್ ಲಿನ್ ಮುಂದಿನ ವಾರ ಭೇಟಿ ನೀಡುತ್ತಿದ್ದಾರೆ ಎಂದು ನೀವು ಕೇಳಿದ್ದೀರಾ
> **Speaker 2:** ಹಾಯ್ ಕಿಮ್, ಹೌದು! ಅವಳು ಮಂಗಳವಾರ ಬೆಳಿಗ್ಗೆ ೧೧ ಗಂಟೆಗೆ AI ನೀತಿಶಾಸ್ತ್ರದ ಕುರಿತು ಭಾಸನ ಮಾಡುತ್ತಿದ್ದ
> **Speaker 1:** ನಾವು ಅವಳ ವಸತಿ ಸೌಕರ್ಯವನ್ನು ಹೊಂದಿಸಿದ್ದೇವೆಯೇ
> **Speaker 2:** ಹೌದು, ಅವಳು ರೆಡ್ಮ್ಯಂಡ್ ಇನ್ ನಲ್ಲಿ ತಂಗಿದ್ದಾಳೆ. ಇದು ಕ್ಯಾಂಪಸ್ಸಿಗೆ ಹತ್ತಿರದಲ್ಲಿದೆ ಮತ್ತು ಶಟಲ್ ಇದೆ.
> **Speaker 1:** ಅದ್ಭುತ. ನಂತರ ನಾವು ಊಟ ಮಾಡುತ್ತಿದ್ದೇವೆಯೇ?
> **Speaker 2:** ಹೌದು, ನಾವು ಗಾರ್ಡನ್ ಕೆಫೆಯಲ್ಲಿ ಟೇಬಲ್ ಬುಕ್ ಮಾಡಿದ್ದೇವೆ. ಇದು ನಿಶಬ್ದವಾಗಿದೆ ಮತ್ತು ಸಸ್ಯಾಹಾರಿ ಆಯ್ಕೆಗಳನ್ನು ಹೊಂದಿದೆ
> **Speaker 1:** ಒಳ್ಳೆಯದು. ವಿಮಾನ ನಿಲ್ದಾಣದಿಂದ ಅವಳನ್ನು ಯಾರು ಕರೆದುಕೊಂಡು ಹೋಗುತ್ತಿದ್ದಾರೆ?
> **Speaker 2:** ನಾನು ಹೋಗುತ್ತಿದ್ದೇನೆ. ಅವಳ ವಿಮಾನ ಸೋಮವಾರ ಸಂಜೆ ೬ ಗಂಟೆಗೆ ಬರುತ್ತಲೇ.
> **Speaker 1:** ಧನ್ಯವಾದಗಳು, ಸ್ಯಾಮ್. ಅವಳಿಗೆ ಅಗತ್ಯವಿರುವ ಎಲ್ಲವನ್ನು ಅವಳು ಹೊಂದಿದ್ದಾಳೆ ಎಂದು ಖಚಿತಪಡಿಸಿಕೊಳ್ಳೋಣ

**Text in Mandarin Chinese**

> **Speaker 1:** 嘿，Sam，你听说林教授下周要来吗?
> **Speaker 2:** 嗨，Kim，听说了！她周二上午 11 点要做一个关于人工智能伦理的演讲。
> **Speaker 1:** 我们安排好她的住宿了吗?
> **Speaker 2:** 是的，她住在雷德蒙德酒店。离园区很近，而且有班车。
> **Speaker 1:** 太好了。我们之后一起吃午饭吗?
> **Speaker 2:** 是的，我们已经在花园咖啡馆订了位子。那里很安静，而且有素食可供选择。
> **Speaker 1:** 太好了。谁去机场接她?
> **Speaker 2:** 是我。她的航班周一下午 6 点到
> **Speaker 1:** 谢谢，Sam。我们需要准备好她需要的一切。

## Project Update

**English translation**

> **Speaker 1:** Hey Nina, how's the rollout of the new productivity feature going?
> **Speaker 2:** Hey; pretty smoothly so far. We've completed internal testing and started the pilot with the marketing team.
> **Speaker 1:** That's great. Any early feedback?
> **Speaker 2:** Actually, yes. They love the interface, but here's the twist—someone discovered it also tracks break times.
> **Speaker 1:** Wait, was that intentional?
> **Speaker 2:** No, it was a side effect of the activity logging. It wasn't in the original spec.
> **Speaker 1:** That could raise privacy concerns.
> **Speaker 2:** Exactly. We've paused the rollout and are working with legal to review the implications.
> **Speaker 1:** Good call. Let's make sure we're transparent with users before we go any further.
> **Speaker 2:** Agreed. I'll draft a communication plan today.

**Comprehension Quiz**

**Temporal Detail (Time-based fact)**   *When did the pilot rollout begin?*

A) **After internal testing** ✓

B) Before internal testing

C) After legal review

D) After the public launch

**Action or Theme (Conceptual understanding)**   *What unexpected issue was discovered during the pilot?*

A) The feature caused system crashes

B) **It tracked break times** ✓

C) It blocked user access

D) It sent automatic emails

**Role or Identity (Semantic reference)**   *Which team was piloting the feature?*

A) Sales

B) Engineering

C) **Marketing** ✓

D) HR

**Dialogue Attribution (Speaker reference)**   *Who raised the concern about privacy?*

A) Speaker on the left side of the screen

B) **Speaker on the right side of the screen** ✓

C) Both

D) None

**Text in Kannada**

**Speaker 1:** ಹೇ ನಿನ, ಹೊಸ ಉತ್ಪಾದಕದ ವೈಶಿಷ್ಟ್ಯದ ಬಿಡುಗಡೆ ಹೇಗೆ ಮುಂದುವರಿಯುತ್ತಿದ?
**Speaker 2:** ಹೇ; ಇಲ್ಲಿಯವರೆಗೆ ಸಾಕಷ್ಟು ಸರಾಗವಾಗಿತ್ತು. ನಾವು ಆಂತರಿಕ ಪರೀಕ್ಷೆಯನ್ನು ಪೂರ್ಣಗೊಳಿಸಿದ್ದೇವೆ ಮತ್ತು ಮಾರ್ಕೆಟಿಂಗ್ ತಂಡದೊಂದಿಗೆ ಪೈಲಟ್ ಅನ್ನು ಪ್ರಾರಂಭಿಸಿದ್ದೇವೆ.
**Speaker 1:** ಅದು ಒಳ್ಳೆಯದು. ಯಾವುದೇ ಆರಂಭಿಕ ಸೂಚನೆ ಸಿಕ್ಕಿದೆಯೇ?
**Speaker 2:** ಹೌದು. ಅವರಿಗೆ ಇಂಟರ್ಫೇಸ್ ಇಷ್ಟವಾಗಿದೆ, ಆದರೆ ಇಲ್ಲಿ ತಿರುವು ಇದೆ - ಯಾರೋ ಇದು ವಿರಾಮದ ಸಮಯವನ್ನು ಸಹ ಟ್ರ್ಯಾಕ್ ಮಾಡುತ್ತೆ ಎಂದು ಕಂಡುಹಿಡಿದಿದ್ದಾರೆ
**Speaker 1:** ನಿರೀಕ್ಷಿಸಿ, ಅದು ಉದ್ದೇಶಪೂರ್ವಕವಾಗಿದೆಯೇ?
**Speaker 2:** ಇಲ್ಲ, ಇದು ಚಟುವಟಿಕೆ ಲಾಗಿಂಗ್‌ನ ಅಡ್ಡಪರಿಣಾಮವಾಗಿತ್ತು. ಅದು ಮೂಲ ವಿಶೇಷಣದಲ್ಲಿ ಇರಲಿಲ್ಲ.
**Speaker 1:** ಅದು ಗೌಪ್ಯತೆ ಕಾಳಜಿಯನ್ನು ಉಂಟುಮಾಡಬಹುದು.
**Speaker 2:** ನಿಖರವಾಗಿ. ನಾವು ಬಿಡುಗಡೆಯನ್ನು ವಿರಾಮಗೊಳಿಸಿದ್ದೇವೆ ಮತ್ತು ಪರಿಣಾಮಗಳನ್ನು ಪರಿಶೀಲಿಸಲು ಕಾನೂನಿಕ ತಂಡದೊಂದಿಗೆ ಕೆಲಸ ಮಾಡುತ್ತಿದ್ದೇವೆ.
**Speaker 1:** ಒಳ್ಳೆಯ ನಿರ್ಧಾರ. ನಾವು ಮುಂದೆ ಹೋಗುವ ಮೊದಲು ಬಳಕೆದಾರರೊಂದಿಗೆ ಪಾರದರ್ಶಕವಾಗಿದ್ದೇವೆ ಎಂದು ಖಚಿತಪಡಿಸಿಕೊಳ್ಳೋಣ.
**Speaker 2:** ಒಪ್ಪುತ್ತೇನೆ. ನಾನು ಇಂದು ಸಂವಹನ ಯೋಜನೆಯನ್ನು ರಚಿಸುತ್ತೇನೆ.

**Text in Mandarin Chinese**

**Speaker 1:** 嘿，Nina，新的生产力功能推出得怎么样？
**Speaker 2:** 嘿，到目前为止还算顺利。我们已经完成了内部测试，并开始与市场部门合作进行前测。
**Speaker 1:** 太棒了。有什么早期反馈吗？
**Speaker 2:** 其实有。他们很喜欢这个界面，但有个小插曲——有人发现它还能记录休息时间。
**Speaker 1:** 等等，这是有意设计的吗？
**Speaker 2:** 不，这是活动记录的副作用。它不在最初的规划中。
**Speaker 1:** 这可能会造成隐私问题。
**Speaker 2:** 没错。我们已经暂停了推出，正在与法务部门合作，调查其影响。
**Speaker 1:** 好的。在进一步推进之前，我们先确保对用户透明。
**Speaker 2:** 同意。我今天会起草一份沟通计划。

## Preparing for a Team Presentation

**English translation**

> **Speaker 1:** Hey Cory, did you get a chance to review the slides for tomorrow's team presentation?
> **Speaker 2:** Yeah, I did. I think slide 5 could use a bit more context.
> **Speaker 1:** I agree. I was thinking of adding a quote from Sophia's report.
> **Speaker 2:** That's a good idea. She had some great insights on user feedback.
> **Speaker 1:** I'll drop it in and send you the updated version by 3pm.
> **Speaker 2:** Perfect. Are we still rehearsing with Christian at 4pm?
> **Speaker 1:** Yep, he booked the small conference room.
> **Speaker 2:** Great. I'll bring a printout for notes.
> **Speaker 1:** Awesome. Let's do it!

**Comprehension Quiz**

**Temporal Detail (Time-based fact)** *What time is the updated slide deck expected to be sent?*

- A) 2 PM
- B) **3 PM** ✓
- C) 4 PM
- D) 5 PM

**Action or Theme (Conceptual understanding)** *What will Christian bring to the rehearsal?*

- A) A laptop
- B) **A printout** ✓
- C) Snacks
- D) A projector

**Role or Identity (Semantic reference)** *Who booked the conference room?*

- A) Marcus
- B) Cory
- C) Sophia
- D) **Christian** ✓

**Dialogue Attribution (Speaker reference)** *Who suggested adding a quote from Sophia's report?*

- A) **Speaker on the left of the screen** ✓
- B) Speaker on the right of the screen
- C) Both
- D) None

**Text in Greek**

**Speaker 1:** Γεια σου Cory, έχεις προλάβει να ελέγξεις τις διαφάνειες για την αυριανή παρουσίαση στην ομάδα;
**Speaker 2:** Ναι, αμέ. Νομίζω ότι η διαφάνεια 5 θα χρειαζόταν λίγες παραπάνω πληροφορίες.
**Speaker 1:** Συμφωνώ. Σκεφτόμουν να προσθέσω ένα απόσπασμα από την αναφορά της Σοφίας.
**Speaker 2:** Πολύ ωραία. Είχε κάποιες καλές ιδέες όσον αφορά τις απαντήσεις από τους χρήστες.
**Speaker 1:** Θα το προσθέσω και θα σου στείλω τις καινούριες διαφάνειες πριν τις 3 το μεσημέρι.
**Speaker 2:** Τέλεια. Η πρόβα στις 4 το απόγευμα ισχύει με τον Christian?
**Speaker 1:** Ναι, έχει κάνει κράτηση τη μικρή αίθουσα συνεδριάσεων.
**Speaker 2:** Ωραία. Θα φέρω μερικές φωτοτυπίες για σημειώσεις.
**Speaker 1:** Τέλεια. Είμαι πανέτοιμη!

**Text in Ukrainian**

**Speaker 1:** Привіт, Корі, у тебе була можливість подивитися слайди для презентації команди взавтра?
**Speaker 2:** Так, встиг. Я думаю що п'ятому слайду не заважало-би трохи більше контексту.
**Speaker 1:** Погоджуюсь. Я як раз думав додати цитату зі звіту Софії.
**Speaker 2:** Це чудова ідея. У неї були добрі інсайти по відгукам користувачів.
**Speaker 1:** Я додам і відправлю тобі оновлену версію до трьох.
**Speaker 2:** Чудово. Ми все ще плануємо практикуватися з Крістіаном о четвертій?
**Speaker 1:** Так, він забронював маленький зал для конференцій.
**Speaker 2:** Чудово. Я принесу роздруківку для заміток.
**Speaker 1:** Супер. Давай!

## Delayed Task

### English translation

**Speaker 1:** Hi Vanya, did you finish the summary for the client call?
**Speaker 2:** Hi Glen, not yet. I got pulled into a last-minute meeting with Greg.
**Speaker 1:** Oh, was that about the budget review?
**Speaker 2:** Yeah, and it ran longer than expected. I'll finish the summary after lunch.
**Speaker 1:** No worries. Just make sure to include the part where Clara asked about timelines.
**Speaker 2:** Got it. I'll also double-check the notes from Sean to make sure I didn't miss anything.
**Speaker 1:** Good call. He usually catches the small stuff.
**Speaker 2:** I'll send it to you and Clara by 2.
**Speaker 1:** Sounds good. Thanks, Vanya.

### Comprehension Quiz

**Temporal Detail (Time-based fact)**   *When will the summary be finished?*

A) Before lunch

B) **After lunch** ✓

C) Tonight

D) Before breakfast

**Action or Theme (Conceptual understanding)**  *Why was Vanya delayed in finishing the summary?*

A) Vanya forgot about it

B) **Vanya had a meeting with Greg** ✓

C) Vanya was out sick

D) Vanya was working on another project

**Role or Identity (Semantic reference)**  *What did Clara ask about in the meeting?*

A) The budget spreadsheet

B) The client's email

C) **The timelines** ✓

D) The project plan

**Dialogue Attribution (Speaker reference)**  *Who will finish the summary?*

A) **Speaker on the left of the screen** ✓

B) Speaker on the right of the screen

C) Both

D) None

**Text in Greek**

> **Speaker 1:** Γεια σου Βάνια, έχεις τελειώσει την περίληψη που θα χρειαστούμε για το τηλεφώνημα με τον πελάτη;
> **Speaker 2:** Γεια σου Γκλεν, όχι ακόμα. Έπρεπε να πάω σε μια συνάντηση της τελευταίας στιγμής με τον Γκρεγκ.
> **Speaker 1:** Α, σχετικά με την αναθεώρηση του προϋπολογισμού;
> **Speaker 2:** Ναι, και τράβηξε πολύ σε διάρκεια. Θα τελειώσω με την περίληψη μετά το μεσημεριανό διάλειμμα.
> **Speaker 1:** Κανένα πρόβλημα. Απλώς μην ξεχάσεις να συμπεριλάβεις τα χρονοδιαγράμματα για τα οποία ρωτούσε η Κλάρα.
> **Speaker 2:** Οκαυ. Επίσης, θα ξαναελέγξω τις σημειώσεις του Σον για να βεβαιωθώ ότι δεν μου ξέφυγε τίποτα.
> **Speaker 1:** Σωστό. Συνήθως είναι καλός με τις λεπτομέρειες.
> **Speaker 2:** Θα το στείλω σε εσένα και την Κλάρα πριν τις 2 το μεσημέρι.
> **Speaker 1:** Καλό μου ακούγεται. Ευχαριστώ, Βάνια.

**Text in Ukrainian**

>**Speaker 1:** Привіт, Ваня, ти закінчив звіт по дзвінку з клієнтом?
>**Speaker 2:** Привіт, Глен, ще ні. Зустріч з Грегом затягнулася в останній момент.
>**Speaker 1:** О, це було з приводу огляду бюджету?
>**Speaker 2:** Так, і вона затягнулася. Я закінчу звіт після обіду.
>**Speaker 1:** Не хвилюйся. Тільки не забудь включити ту частину, де Клара питає про строки.
>**Speaker 2:** Зрозумів. Я також перевірю записи Шона, щоб впевнитися, що я нічого не упустив.
>**Speaker 1:** Це добра ідея. Зазвичай він помічає дрібниці.
>**Speaker 2:** Я відправлю це тобі і Кларі до двох.
>**Speaker 1:** Чудово. Дякую, Ваня.

## Planning of a Lunch Outing

**English translation**

>**Speaker 1:** Hey Ollie, want to grab lunch with me and Joshua today?
>**Speaker 2:** Sure! Where are you guys thinking of going?
>**Speaker 1:** Joshua suggested that new Mediterranean place near the library.
>**Speaker 2:** Oh, I've heard good things. Do they have vegetarian options?
>**Speaker 1:** Yeah, I checked the menu. They've got falafel wraps and a lentil bowl.
>**Speaker 2:** Nice. What time are you heading out?
>**Speaker 1:** Around 12:15pm. We're meeting Justin at the lobby.
>**Speaker 2:** Cool, I'll finish up this email and meet you there.
>**Speaker 1:** Awesome. I'll text you if anything changes.

**Comprehension Quiz**

**Temporal Detail (Time-based fact)**   *What time are they planning to head out for lunch?*

- A) 12:00 PM
- B) **12:15 PM** ✓
- C) 12:30 PM
- D) 12:45 PM

**Action or Theme (Conceptual understanding)**   *What kind of food does the restaurant serve?*

- A) Italian
- B) **Mediterranean** ✓
- C) Thai
- D) Mexican

**Role or Identity (Semantic reference)**   *Who suggested the restaurant?*

- A) Daniel
- B) Ollie
- C) **Joshua** ✓
- D) Justin

**Dialogue Attribution (Speaker reference)**  *Who checked whether the restaurant has vegetarian options?*

- A) **Speaker on the left of the screen** ✓
- B) Speaker on the right of the screen
- C) Both
- D) None

**Text in Greek**

> **Speaker 1:** Γεια σου Όλι, θα έρθεις για μεσημεριανό με μένα και τον Τζόσουα σήμερα;
> **Speaker 2:** Ναι αμέ! Πού σκέφτεστε να πάτε;
> **Speaker 1:** Ο Τζόσουα πρότεινε αυτό το νέο μεσογειακό εστιατόριο κοντά στη βιβλιοθήκη.
> **Speaker 2:** Α, έχω ακούσει καλά λόγια. Έχουν επιλογές για χορτοφάγους;
> **Speaker 1:** Ναι, έλεγξα το μενού. Έχουν φαλάφελ και φακιές σούπα.
> **Speaker 2:** Ωραία. Τι ώρα θα ξεκινήσεις;
> **Speaker 1:** Γύρω στις 12:15 μ.μ. Θα συναντήσουμε τον Τζάστιν στο λόμπι.
> **Speaker 2:** Ωραία, θα τελειώσω αυτό το email και θα σε συναντήσω εκεί.
> **Speaker 1:** Τέλεια. Θα στείλω μήνυμα αν αλλάξει κάτι.

**Text in Ukrainian**

> **Speaker 1:** Привіт, Оллі, хочеш сьогодні пообідати зі мною і Джошуа?
> **Speaker 2:** Звісно! Куди ви, хлопці, збираєтесь піти?
> **Speaker 1:** Джошуа запропонував новий середземноморський ресторанчик поруч з бібліотекою.
> **Speaker 2:** О, я чув добрі відгуки. Чи є у них вегетеріанські блюда?
> **Speaker 1:** Так, я подивився меню. У них є фалафель в ролах і суп з сочевиці.
> **Speaker 2:** Супер. Коли ви виходите?
> **Speaker 1:** Близько 12:15. Ми зустрічаємося з Джастіном у холі.
> **Speaker 2:** Класно, я закінчу тут один електронний лист і зустрінусь з вами там.
> **Speaker 1:** Чудово. Я напишу тобі, якщо щось зміниться.

**Task Clarification**

**English translation**

> **Speaker 1:** Hey Reece, did you understand what Maria meant in her message about the video files?
> **Speaker 2:** Sort of. She said we need to label them, but I wasn't totally sure which ones she meant.
> **Speaker 1:** I think she was talking about the ones from our safari trip last week.
> **Speaker 2:** Oh, the ones where we saw the lions near the river and the elephants crossing the road?
> **Speaker 1:** Exactly Reece. Maria wants us to label each clip based on the animals that appear in them.
> **Speaker 2:** That makes sense. Should we double-check with Rita just to be sure?
> **Speaker 1:** Yeah, I'll message her now. I'd rather confirm than guess wrong.

**Speaker 2:** Cool. I'll start sorting the files into folders while you check.
**Speaker 1:** Perfect. Let's try to finish labeling everything by tomorrow afternoon.

**Comprehension Quiz**

**Temporal Detail (Time-based fact)**   *When do they plan to finish labeling the video files?*

   A) **By tomorrow afternoon** ✓

   B) By the end of the week

   C) Before lunch today

   D) Next Monday

**Action or Theme (Conceptual understanding)**   *What is the task they are discussing?*

   A) Editing a video

   B) **Labeling video files** ✓

   C) Writing a report

   D) Scheduling a meeting

**Role or Identity (Semantic reference)**   *What animals appeared in the video files?*

   A) Dolphins and sharks

   B) Elephants and giraffes

   C) **Lions and elephants** ✓

   D) Horses and dolphins

**Dialogue Attribution (Speaker reference)**   *Who will check with Rita for clarification?*

   A) Speaker on the left of the screen

   B) **Speaker on the right of the screen** ✓

   C) Both

   D) None

**Text in Greek**

>**Speaker 1:** Γεια σου Reece, κατάλαβες τι ακριβώς εννοούσε η Maria στο μήνυμά της, για τα αρχεία βίντεο;
>**Speaker 2:** Στο περίπου. Είπε ότι πρέπει να τα ονομάσουμε, αλλά δεν ήμουν απόλυτα σίγουρη σε ποια ακριβώς αναφερόταν.
>**Speaker 1:** Νομίζω ότι μιλούσε για αυτά από το σαφάρι που πήγαμε την περασμένη εβδομάδα.
>**Speaker 2:** Α, αυτά όπου είδαμε τα λιοντάρια κοντά στο ποτάμι και τους ελέφαντες να διασχίζουν τον δρόμο;
>**Speaker 1:** Ακριβώς Reece. Η Maria θέλει να ονομάσουμε κάθε κλιπ με βάση τα ζώα που εμφανίζονται σε αυτά.

**Speaker 2:** Ναι σωστό μου ακούγεται. Να ξαναρωτήσουμε τη Rita για να είμαστε σίγουροι?
**Speaker 1:** Ναι, θα της στείλω μήνυμα τώρα. Καλύτερα να επιβεβαιώσω παρά να κάνουμε λάθος.
**Speaker 2:** Ωραία. Θα αρχίσω να ταξινομώ τα αρχεία σε φακέλους όσο εσύ ελέγχεις.
**Speaker 1:** Τέλεια. Ελπίζω να τελειώσουμε με την ονομασία αρχείων μέχρι αύριο το απόγευμα.

**Text in Ukrainian**

**Speaker 1:** Привіт, Ріс, ти зрозумів, що Марія мала на увазі у своєму повідомленні про відеофайли?
**Speaker 2:** Наче так. Вона сказала, що нам потрібно їх підписати, але я не зовсім зрозумів, які саме.
**Speaker 1:** Здається, вона казала про відео з нашого сафарі минулого тижня.
**Speaker 2:** О, це там, де ми зняли левів біля річки і слонів, які переходили дорогу?
**Speaker 1:** Саме так, Ріс. Марія хоче, щоби ми підписали кожен кліп в залежності від тих тварин, котрі в ньому з'являються.
**Speaker 2:** Логічно. Може ще раз перевіримо у Ріти, щоби впевнитися?
**Speaker 1:** Так, я їй зараз напишу. Краще перевірити, а не помилитися.
**Speaker 2:** Чудово. Я почну сортувати файли по папках, поки ти перевіряєш.
**Speaker 1:** Чудово. Давай постараємося закінчити підписувати все до завтрашнього вечора.

# 2 Questionnaire

## Satisfaction Survey

**S$_1$** *How easy was it to understand the conversation you just heard?*

| 1 | 2 | 3 | 4 | 5 | 6 | 7 |
|---|---|---|---|---|---|---|
| Very difficult | Difficult | Somewhat difficult | Neutral | Somewhat easy | Easy | Very easy |

**S$_2$** *How clearly could you distinguish between the two speakers?*

| 1 | 2 | 3 | 4 | 5 | 6 | 7 |
|---|---|---|---|---|---|---|
| Not clearly at all | Slightly clearly | Somewhat clearly | Neutral | Clearly | Very clearly | Extremely clearly |

**S$_3$** *How immersive and engaging did the translated audio feel?*

| 1 | 2 | 3 | 4 | 5 | 6 | 7 |
|---|---|---|---|---|---|---|
| Not immersive or engaging at all | Slightly immersive / engaging | Somewhat immersive / engaging | Neutral | Moderately immersive / engaging | Very immersive / engaging | Extremely immersive and engaging |

**S$_4$** *How distracting, if at all, was the original audio?*

| 1 | 2 | 3 | 4 | 5 | 6 | 7 |
|---|---|---|---|---|---|---|
| Not at all distracting | Slightly distracting | Somewhat distracting | Neutral | Distracting | Very distracting | Extremely distracting |

**S₅** *How satisfied were you with the overall listening experience?*

| 1 | 2 | 3 | 4 | 5 | 6 | 7 |
|---|---|---|---|---|---|---|
| Very dissatisfied | Dissatisfied | Somewhat dissatisfied | Neutral | Somewhat satisfied | Satisfied | Very satisfied |

## Workload Survey

**TLX₁ Mental demand**: *How mentally demanding was the task?*

| 1 | 2 | 3 | 4 | 5 | 6 | 7 |
|---|---|---|---|---|---|---|
| Very low | Low | Slightly low | Neutral | Slightly high | High | Very high |

**TLX₂ Temporal demand**: *How hurried or rushed did you feel during the task?*

| 1 | 2 | 3 | 4 | 5 | 6 | 7 |
|---|---|---|---|---|---|---|
| Not at all rushed | Slightly rushed | Somewhat rushed | Neutral | Moderately rushed | Very rushed | Extremely rushed |

**TLX₃ Effort**: *How hard did you have to work to understand the conversation?*

| 1 | 2 | 3 | 4 | 5 | 6 | 7 |
|---|---|---|---|---|---|---|
| Very low effort | Low effort | Slightly low effort | Neutral | Slightly high effort | High effort | Very high effort |

**TLX₄ Performance**: *How confident are you that you understood the key points of the conversation?*

| 1 | 2 | 3 | 4 | 5 | 6 | 7 |
|---|---|---|---|---|---|---|
| Not at all confident | Slightly confident | Somewhat confident | Neutral | Confident | Very confident | Extremely confident |

## Recap Survey

**R₁** *Overall, which condition did you find most helpful for understanding the conversation? Why? (You can refer to the listening sessions by their name, e.g., "Session 1", "Session 2", etc.).*

**R₂** *How important was it for you to distinguish between the two speakers? Did any condition make that easier or harder?*

**R₃** *Where there any conditions that felt confusing, overwhelming, or unnatural? What stood out to you?*

**R₄** *Do you have any suggestions for improving the audio translation experience in multilingual meetings?*